\documentclass{aa}

\usepackage{hyperref}
\usepackage{graphicx}
\def\Cloudy{{\sc Cloudy}}
\usepackage{datetime}
\newdateformat{monthyeardate}{\monthname[\THEMONTH] \THEDAY, \THEYEAR}
\newcommand{\orcidlogo}[2][9pt]{%
  \href{https://orcid.org/#2}{%
    \includegraphics[width=#1]{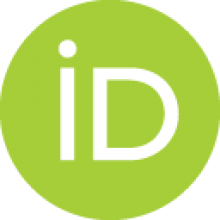}
  }%
}
\usepackage{txfonts}
\begin{document} 

   \title{New Insights with XRISM \& Cloudy:\\ A novel Column Density Diagnostic}


   \author{Chamani M. Gunasekera\orcidlogo{0000-0002-4634-5966}
          \inst{1}
          \and
          Peter A. M. van Hoof\orcidlogo{0000-0001-7490-0739}\inst{2}
          \and
          Masahiro Tsujimoto\orcidlogo{0000-0002-9184-5556}\inst{3}
          \and
          Gary J. Ferland\orcidlogo{0000-0003-4503-6333}\inst{4}
          }

   \institute{Space Telescope Science Institute, 3700 San Martin Drive, Baltimore, MD 21218, USA
   \and 
   Royal Observatory of Belgium, Ringlaan 3, B-1180 Brussels, Belgium
   \and
   Institute of Space and Astronautical Science (ISAS), Japan Aerospace Exploration Agency (JAXA), 3-1-1 Yoshinodai, Chuo-ku, Sagamihara, Kanagawa 252-5210, Japan
   \and
   University of Kentucky, 506 Library Drive, Lexington, KY 40506, USA
   \\
             \email{cgunasekera@stsci.edu}
             }

   \date{Received November 22, 2024 / Accepted January 18, 2024}

 
  \abstract
   {}
   {We present a simple, yet powerful column density diagnostic for plasmas enabled by X-ray microcalorimeter observations.}
   {With the recent developments of the spectral simulation code \Cloudy, inspired by the high spectral resolution of the X-Ray Imaging and Spectroscopy Mission (XRISM) and the Advanced Telescope for High Energy Astrophysics (Athena), we make predictions for the intensity ratio of the resolved fine-structure lines Ly$\alpha_1$ and Ly$\alpha_2$ of H-like ions.}
   {We show that this ratio can be observationally constrained and used as a plasma column density indicator. We demonstrate this with a XRISM observation of the high-mass X-ray binary Centaurus X-3.}
   {This diagnostic is useful for a wide range of X-ray emitting plasmas, either collisionally or radiatively ionized.}


   \maketitle
%

\section{Introduction} 
\label{sec:intro} 

Spectral resolution is a primary limiting factor in studying the Universe's hottest
phenomena. The X-Ray Imaging and Spectroscopy Mission (XRISM; \citealt{tashiro2020})
launched in 2023 is poised to unlock new insights into the fundamental nature of the
X-ray universe, with its state-of-the-art {Resolve} microcalorimeter spectrograph
\citep{ishisaki2022}. However, we will need new tools to fully use the XRISM data
as it will be able to resolve many spectral features for the first time. One of them
is the fine-structure doublet of the Lyman $\alpha$ line of H-like ions; i.e., the
2p~$^{2}P_{3/2}$ $\rightarrow$ 1s~$^{2}S_{1/2}$ and 2p~$^{2}P_{1/2}$ $\rightarrow$ 1s~$^{2}S_{1/2}$
transitions. We hereafter call them Ly$\alpha_1$ and Ly$\alpha_2$, respectively, and
discuss their line intensity ratio.

\Cloudy{}\ (last reviewed by \citealt{2023RMxAA..59..327C, 2023RNAAS...7..246G}) is one of the most widely used spectral simulation codes, for plasmas in non-local thermodynamic equilibrium.
\Cloudy{} handles one- and two-electron systems with a consistent method along
iso-sequences. The two-electron iso-sequence with optical emission lines was expanded in
\citet{2012MNRAS.425L..28P, 2013MNRAS.433L..89P}, while the code uses atomic databases
for the many-electron systems \citep{2015ApJ...807..118L, 2022Astro...1..255G}. Recent
work, described in \cite{2024arXiv241201606G}, has expanded the one-electron
systems to simulate X-ray spectra matching the resolution of current and future
microcalorimeter observations. The detailed microphysics required to resolve the Lyman
series doublet, at the resolution of XRISM, is presented in \cite{2024arXiv241201606G}. This development will be the primary component of the \Cloudy{} 2025 release.


\begin{figure}[h]
 \centering
 \includegraphics[width=0.78\columnwidth]{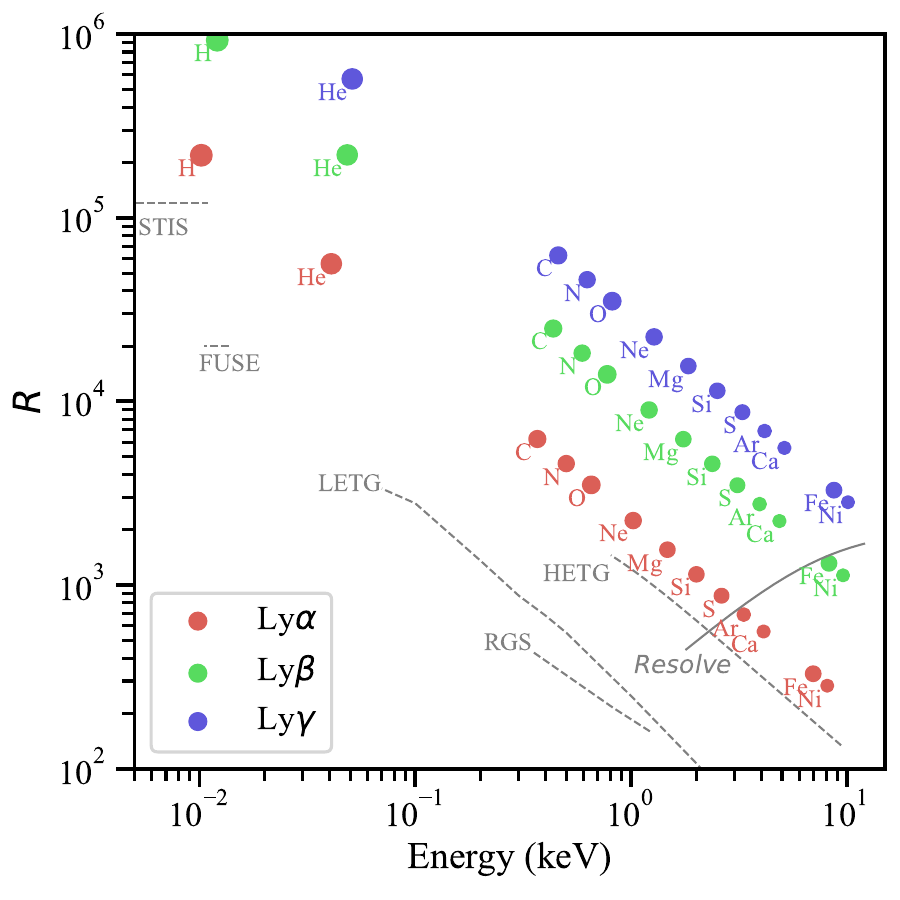}
 \caption{Required resolving power ($R \equiv E/\Delta E$) to separate the
 Ly$\alpha_{1,2}$ (red), $\beta_{1,2}$ (green), and $\gamma_{1,2}$ (blue) fine-structure
 doublets of major elements compared to the instrumental resolving power of
 Hubble Space Telescope Imaging Spectrograph (STIS) \citep{kimble1998}, Far Ultraviolet Spectroscopic Explorer (FUSE) \citep{moos2000}, Chandra LETG
 \citep{brinkman2000a} and HETG \citep{canizares2005}, {XMM-Newton} RGS
 \citep{denherder2001}, and XRISM Resolve \citep{ishisaki2022} using the full-width half maximum (FWHM). The
 elements below the curves can be resolved with the spectrometer. The symbol size
 represents the logarithm of the solar abundance of the elements \citep{anders1989}.}
 \label{f01}
\end{figure}

The resolving power ($R$) required to separate the Lyman series doublet is shown for
abundant elements in Figure~\ref{f01}. The energy split of fine-structure levels stems from the spin-orbit coupling term of the Hamiltonian, which increases as a function of
the atomic number ($Z$) at a rate $\Delta E \propto Z^{4}$. 
As detailed in \cite{2024arXiv241201606G} the required resolving power $R = E/\Delta E \propto Z^{-2}$ becomes less demanding for
higher $Z$ elements.  All high-resolution X-ray spectrometers before XRISM --- the Chandra Low and 
the High Energy Transmission Gratings  (LETG \citealt{brinkman2000a}; HETG \citealt{canizares2005}) and the Reflection Grating Spectrometer (RGS)
\citep{denherder2001} onboard the X-ray Multi-Mirror Mission Newton (XMM-Newton)--- are based on dispersion
spectrometry, which has a spectral resolution approximately constant in wavelengths
$\Delta\lambda$, thus $R = \lambda/\Delta\lambda \propto E^{-1} \propto Z^{-2}$. The
resolution of the dispersion spectrometers and that required to resolve fine-structure
levels have the same dependence on $E$ by coincidence. As a result, none of the Lyman
series doublets for any $Z$ was resolved by these spectrometers.

The X-ray microcalorimeter onboard XRISM uses an entirely different spectroscopic
technique based on non-dispersive X-ray microcalorimetery, whose resolution is
approximately constant in energy $\Delta E$, thus $R \propto E^{1}$. For $Z \ge 18$
(Ar), the Ly$\alpha$ doublet is resolved with XRISM for the first time for any
astronomical source at a redshift of approximately zero, except for the Sun. It can even
resolve the Ly$\beta$ doublet for $Z \ge 26$ (Fe). This is also the case for other
fine-structure levels such as He-like ions' $x$ and $y$ lines of the He$\alpha$ ($n=2
\rightarrow 1$) transitions. For He-like ions, we denote $w$, $x$, $y$, and $z$,
respectively for the $^{1}P_{1} \rightarrow{} ^{1}S_{0}$, $^{3}P_{2} \rightarrow{}
^{1}S_{0}$, $^{3}P_{1} \rightarrow{} ^{1}S_{0}$, and $^{3}S_{1} \rightarrow{} ^{1}S_{0}$
transitions from the 1s2p or 1s2s levels to the 1s$^{2}$ ground state.


In this Letter, we focus on the Ly$\alpha$ doublet. We will show that the intensity
ratio of the doublet is a useful diagnostic of the physical conditions of X-ray emitting
plasmas. We present the theoretical predictions made by \Cloudy{} in Section~\ref{sec:cloudy}
and apply the method to actual observational data with XRISM in Section~\ref{sec:XRISM}.

\section{Spectral Simulations Using Cloudy}
\label{sec:cloudy}

\begin{figure}
 \centering
 \includegraphics[width=0.78\columnwidth]{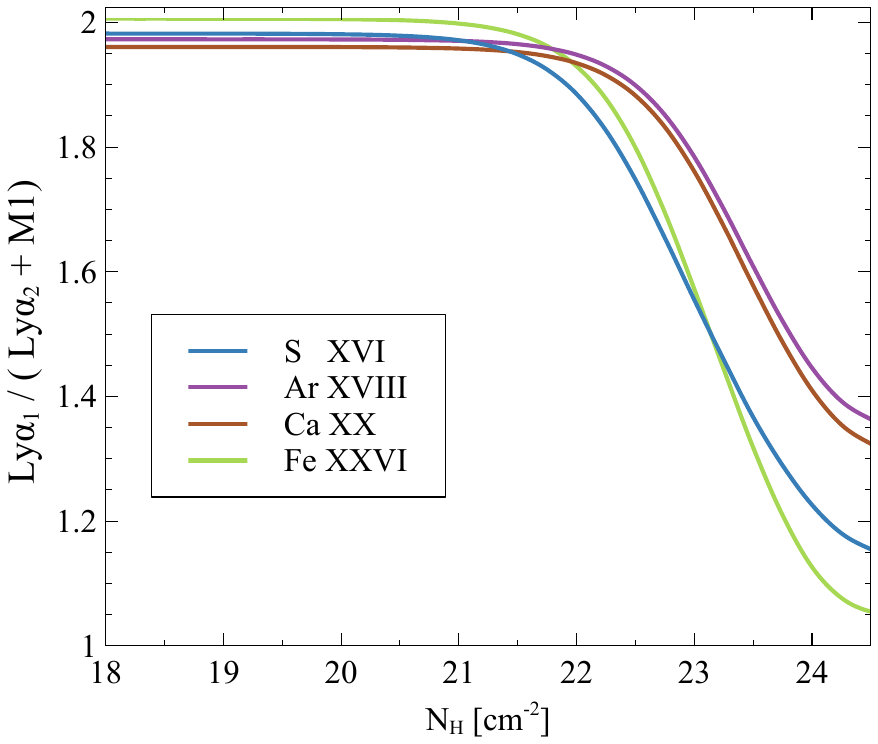}
 \caption{Line intensity ratio of the Ly$\alpha$ doublet for selected H-like ions as a function of the plasma column density calculated with \Cloudy{}. Note that the absolute value of $N_{\mathrm{H}}$ changes for different plasma environments, as the line optical depth is converted into $N_{\mathrm{H}}$ using the chemical abundance, charge and level populations of the simulation. \Cloudy{} provides Ly$\alpha_2$ + M1 as a "blend" for Fe\,{\sc XXVI} only. For the other one-electron species, these line must be obtained separately in \Cloudy{} and summed by hand.}
 \label{fig:KLineRatio}
\end{figure}

We present a \Cloudy{} simulation for the plasma at the Perseus Cluster core
\citep{hitomi16}. Being one of the brightest galaxy clusters observed in X-rays, it
provides an ideal case to explore the physics. The model assumes a collisionally-ionized
plasma at a constant temperature of 4.7$\times10^7$ K and hydrogen density of
$10^{-1.5}$~cm$^{-3}$ with varying plasma thickness in the line of sight evaluated as
the H-equivalent column density ($N_{\mathrm{H}}$). We also include a microturbulent
velocity of 150 km s$^{-1}$, ensuring more realistic line shielding and pumping. 

The line radiative transport is handled using a unified approach with the
goal of having a single class (a group of objects in the code that share the same properties and behaviours) in \Cloudy{} with the appropriate atomic parameters.
Line transfer is handled within this framework. 
As summarized by \citet{1962MNRAS.125...21H},
four cases can be identified depending on the lifetimes of the upper and
lower states and the importance of Doppler and radiative broadening.
\citet{1987cup..bookR....K} gives further details.
By default, \Cloudy{} assumes partial redistribution for resonance lines 
and complete redistribution for subordinate lines.
These assumptions can be changed with options in the user interface
although tests show that these do not influence the conclusions presented here.

Figure~\ref{fig:KLineRatio} shows how the intensity ratio of the Ly$\alpha$ doublet
changes as $N_{\mathrm{H}}$ increases for some selected elements. Two simple limits are
apparent. At low column densities, where the line optical depth is small, the line
photons escape freely without interactions. 
\Cloudy\ Ly$\alpha_j$ intensities are proportional
to the populations of the upper levels, which are assumed to be proportional to their
statistical weights (further discussed in \citealt{2024arXiv241201606G}).
Hence the ratio is predicted to be $2:1$ with the Ly$\alpha_1$ line being
stronger. This is close to the value observed in the \ion{Fe}{XXVI} doublet in the corona X-rays in the
Sun, in which the lines are optically thin \citep{tanaka1986}.

As the optical depth increases, line photons undergo an increasing number of absorptions
and re-emissions before escaping.  As a result, the photons we observe at Earth will
have the physical conditions imprinted on them where they were last
re-emitted. In other words, the line flux will be determined by the physical conditions
at that point.  More quantitatively, this can be understood by looking at the
Eddington-Barbier approximation, which states that the emergent flux is determined by
the source function at the location where the line optical depth $\tau$ reaches $2/3$
when integrated from the observers point of view \citep{2003rtsa.book.....R}.
Assuming the conditions are sufficiently similar in the regions where the Ly$\alpha_1$
and Ly$\alpha_2$ lines reach $\tau = 2/3$, the Ly$\alpha_1$/Ly$\alpha_2$ line ratio will
approach 1:1 for high column densities.  Note that \Cloudy\ does not code the
Eddington-Barbier approximation explicitly, but rather uses the escape probability
formalism to compute the line fluxes. This should, however, give very similar results.
This implies that the Ly$\alpha$ doublet intensity ratio is sensitive to column
densities in the range of $\approx 10^{20}-10^{24}$~cm$^{-2}$ with different ranges for
different elements, which is the appropriate range observed in many systems.


\section{Observational Demonstration using XRISM}
\label{sec:XRISM} 
We demonstrate the proposed diagnostic using actual observation data with XRISM. We take
Centaurus X-3 (Cen X-3) as an example. Cen X-3 is an eclipsing high-mass X-ray binary
comprised of a neutron star (NS) and an O6.5 II-III star \citep{schreier1972}. The X-ray
spectra during eclipses are not contaminated by the direct emission from the NS,
upon which absorption features are imprinted. Therefore, it is more suitable for the
emission line ratio diagnostics focusing on the diffuse emission from the photoionized plasma.

\begin{figure}
 \centering
 \includegraphics[width=0.9\columnwidth]{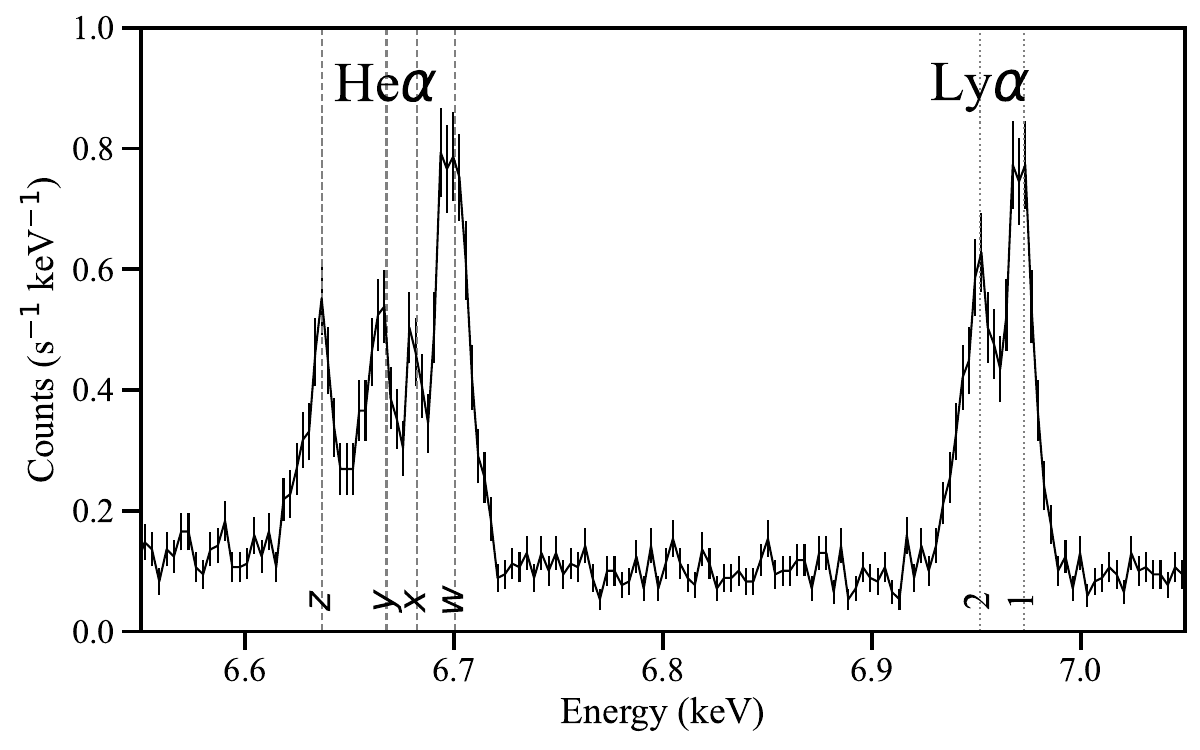}
 \caption{{Resolve} spectrum of Cen X-3 during eclipse \citep{pradhan2024a} of
 \ion{Fe}{XXV} He$\alpha$ and \ion{Fe}{XXVI} Ly$\alpha$ complex. The fine-structure levels of the
 complexes are clearly resolved.}
 \label{f02}
\end{figure}

Cen X-3 was observed with {Resolve} on February 12--15, 2024 (sequence number
300003010; \citealt{mochizuki2024}). Two eclipses were covered during the XRISM
observation. We integrated the X-ray spectrum over the eclipses for 48~ks
(Figure~\ref{f02}). The fine-structure levels in the He$\alpha$ and Ly$\alpha$ complexes
are indeed resolved.
The Ly$\alpha$ line complex was fitted using two Gaussian lines plus a power law model
to account for the underlying continuum. The line centers were fixed to the values in
the APED database \citep{smith01} but they were allowed to shift collectively
to account for possible systematic velocities and energy gain calibration
uncertainty. The line intensities were fitted individually. The widths of the two lines
(1 and 2) were assumed to be the same and they were fitted simultaneously. A successful
fit was obtained with a reduced $\chi^2 < 1.2$ and the line ratio was derived as 1.35 $\pm$ 0.11.

\begin{figure}
 \vspace*{3mm}
 \centering
 \includegraphics[width=0.7\columnwidth]{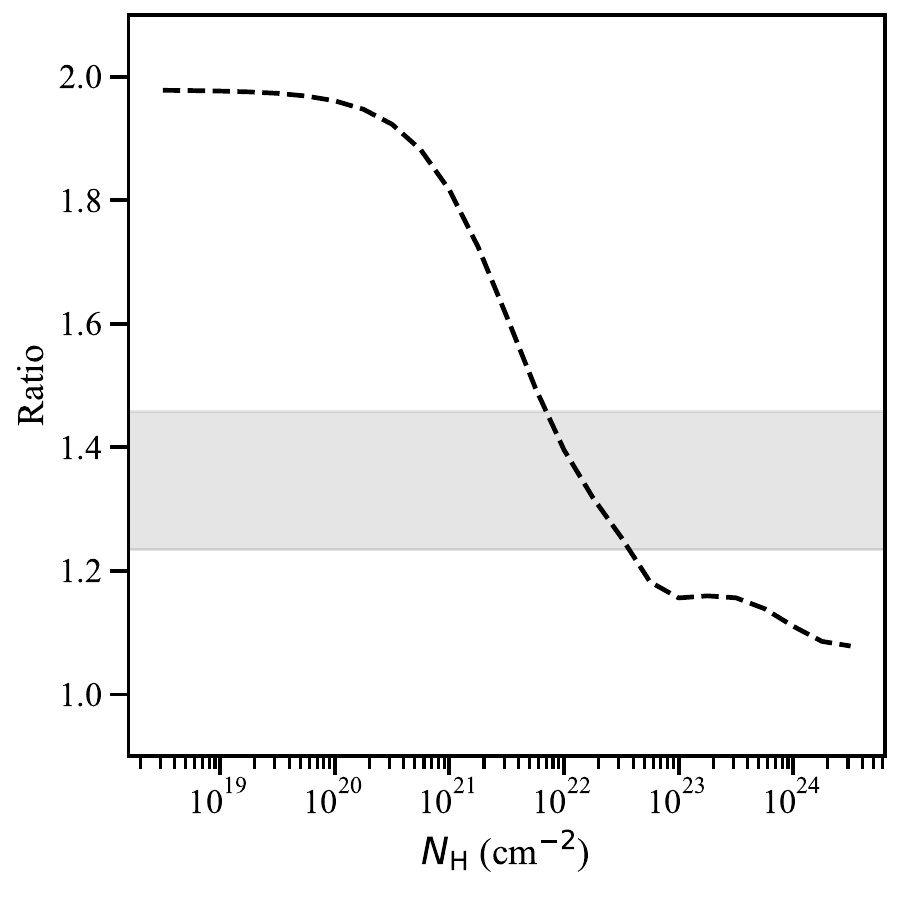}
 \caption{Ly$\alpha_1$/Ly$\alpha_2$ ratio calculated with \Cloudy{} for the Cen X-3
 setting (dashed curve) and the observed ratio (shaded area). Though minor for this
 setup, the M1 transition (2s)$^{2}S_{1/2}$ $\rightarrow$ (1s)$^{2} S_{1/2}$ intensity
 is added to Ly$\alpha_2$, which cannot be resolved with {Resolve}.}
 \label{f03}
\end{figure}

We now apply the diagnostics proposed in Section~\ref{sec:cloudy}. We ran a \Cloudy{}
simulation customized for Cen X-3. The photo-ionizing emission is provided by the NS,
which is represented by a power law of a photon index of --1.8 and a luminosity
$L_{\mathrm{X}} =10^{37}$~erg~s$^{-1}$ in the 1--1000 Ryd range as derived from the
data. The emission lines from the photo-ionized plasma are observed when the NS is
eclipsed by the O star, indicating that the plasma size is no smaller than the radius of
the O star (12~$R_{\mathrm{\odot}}$). We thus set the inner radius
$r_{\mathrm{in}}=10^{12}$~cm. The ionization parameter $\log{\xi} \sim 4$ from the line
intensity ratio of H-like versus He-like ions. These yield the electron density $n =
L_{\mathrm{X}} / (r_{\mathrm{in}}^{2}\xi) =10^9$~cm$^{-3}$. We assumed a plane-parallel
geometry. The turbulent velocity was set to zero. Figure~\ref{f03} shows the result. We
constrain the hydrogen-equivalent column density of the plasma to be $\sim 2 \times
10^{22}$~cm$^{-2}$. This is consistent with other methods such as the He$\alpha$ $z$/$w$
ratio \citep{chakraborty2021} within a factor of a few, verifying the validity of this
diagnostic.

\section{Summary}
The present study highlights the groundbreaking potential for new science that can be
uncovered by using the latest \Cloudy\ developments to understand microcalorimeter
observations. The column thickness in the line of sight is a crucial quantifier of the
X-ray emitting plasma in all systems. We presented a novel column density diagnostic
using the intensity ratio of the resolved Ly$\alpha$ doublet of one-electron atoms,
which changes from 2 to 1 as the optical thickness of the line center increases as
$N_{\mathrm{H}}$ increases. Different elements have sensitivity in different ranges of
the $N_{\mathrm{H}}$ for their differences in abundance, charge and level populations,
oscillator strength, line profiles, and so on, which can be calculated or taken into
account in \Cloudy{} simulations. We demonstrated this for Cen X-3 using the \ion{Fe}{XXVI}
Ly$\alpha$ doublet resolved with XRISM data and the \Cloudy{} simulation and constrained
its $N_{\mathrm{H}}$ value.

\section*{Data availability}
All \Cloudy{} models and subsequent figures used in the present paper are available
at \url{https://gitlab.nublado.org/cloudy/papers/-/tree/main/arXiv.2411.15357}.

\begin{acknowledgements}
CMG and GJF were supported by JWST-AR-06419 and JWST-AR-06428. This research made use of the JAXA's high-performance computing system JSS3.
\end{acknowledgements}

%
%
\bibliographystyle{aa}
\bibliography{main}
\end{document}